# Incorporating the Environmental Dimension into Multidimensional Poverty Measurement: An Initial Proposition

**Pérez-Cirera, V. López-Corona, O. Teruel, G. Carrera, F. Reyes, M. García, A.**

*Institute for Equitable Development (EQUIDE). Universidad Iberoamericana, Ciudad de México. Prolongación Paseo de la Reforma # 880, Lomas de Santa Fé, C.P. 01219, Ciudad de México.*

**ABSTRACT**

Multidimensional poverty measurement has captured the attention of policy-makers and researchers during recent years. Mexico is one of the most advanced countries in the measurement of poverty beyond income indicators. However, both in Mexico and other countries which have attempted at measuring poverty from a multidimensional perspective, the environmental dimension, has been well under-represented. Based on international evidence and, using the welfare-rights based methodological framework used in Mexico to measure multi-dimensional poverty officially, the paper proposes an indicator and six sub-indicators for measuring the lack of a minimum welfare to fulfil the right to a healthy environment through 6 sub-indicators measuring effective access, quality and continuity with regards to water, energy, biodiversity, air, spatial health, waste management and the vulnerability to poverty due climate change impacts.

**INTRODUCTION**

During the past twenty years, academically, poverty measurement and poverty dynamics have been a permanent concern as generally decreasing but still prevalent and large social problem. In the case of Latin American countries, the crisis in the eighties that translated in the so called lost decade, not only in terms of economic growth but also in terms of growing inequality spurred an increased interest in the study of poverty measurement and poverty dynamics from researchers from the region and elsewhere. The current environmental and dragging economic crisis globally forces scientists and policy-makers to craft conceptual frameworks and methods to identify who is likely to become poor and how can we become more resilient to poverty. The response to these questions demands a deeper and more integral understanding of poverty.

The traditional approach to the study of poverty has used income as the sole proxy of welfare. The limitations of this approach have been questioned for several decades now. Beginning in the sixties, the Movement of Social Indicators proposed that income was a limited and insufficient indicator to evaluate social welfare (Andrews and Withey, 1976; Andrews et al, 1989; Vogel, 1985). This movement defended the use and construction of additional indicators in order to evaluate welfare such as the amount of money spent in health, doctors per inhabitant, literacy rate, life expectancy, disease incidence, child mortality, birth indicators, etc. (Rojas, 2008). By the early nineties, several approaches urging to take a multidimensionality approach started to multiply both from a needs-based perspective (Altimir, 1979; Boltvinik, 1992) as well as from a rights-based perspective (UNDP, 1997).

Mexico is an emblematic case in Latin America and the world, taking the multidimensional approach to an official poverty measurement methodology. Other emblematic cases which use multidimensional perspectives are Chile and Colombia.

In the official measurement of multidimensional poverty undertaken by the National Commission for the Evaluation of Social Policy (CONEVAL, for its acronym in Spanish), some specific environmental aspects are included for example, the access to running water or the primary use of biofuels for domestic heat needs in the indicator measuring the right to a worthy home. However, environmental elements which have critical impacts on welfare are missing.

From a welfare perspective, it has become evident that the quality of the environment and the health of ecosystems are critical to human well-being (Costanza et al. 1997; Millennium Ecosystem Assessment, 2005). From a human rights perspective, considerations in regards to the environment were considered quite provocative still a decade ago. Nowadays, the right to a healthy environment is part of the constitutions of most developed countries and rapidly being incorporated in developing countries. In correspondence with the emerging Economic, Social, Cultural and Environmental human rights international framework DESCA, for its acronym in Spanish, in 2011, Mexico made a constitutional reform which, amongst other, included the right to a healthy environment, as part of Mexico´s citizens' social rights. In correspondence, the General Law on Social Development (LGDS, for its acronym in Spanish), enacted in Mexico in 2004,  mandates the consideration of all social rights considered in the constitution, including the right to a healthy environment, within the official multidimensional poverty measurement. CONEVAL has started to engage in an academic process to define what the right to a healthy environment could mean for any Mexican citizen, with an initial interesting proposition, considering regulatory quality thresholds for some environmental elements: air, water and soil (see Muñoz, 2015). However, up to now, there is no proposition on how to include this right from a minimum-welfare perspective, useful for poverty measurement. This paper proposes a conceptual framework and an initial set of indicators to include the environmental dimension within Mexico's official multidimensional poverty measurement methodology and as a basis to spur discussions for its consideration elsewhere.

**1.      Multidimensional poverty measurement in Mexico**

In 2004, the LGDS established a set of criteria to CONEVAL for the measurement of poverty based on the lack of fulfilment of social rights. Between 2006 and 2009, CONEVAL developed two lines of research. The first oriented at defining the theoretical framework for the multidimensional measurement of poverty and the second, to determine the information requirements. As a result, CONEVAL developed two work-programs (i) economic welfare and, (ii) social rights and, a set of ground-rules:

1. Shortcomings are to be evaluated through binary variables
2. All rights need to be weighted similarly
3. The lack of fulfilment to any social right is considered an absolute  shortcoming in

the welfare of a household

From this, an "Index of Social Privation", which goes from 0 to 1 was derived.  For economic welfare, CONEVAL uses a standard monetary measure in which per capita income is compared to the acquisition of two bundles. A food bundle and another bundle which adds other usually consumed goods.  Thresholds in the realms of social rights are analogous to poverty lines, making it possible to lineally sum-up rights up to a point (above

3) in which a person is extremely "lack-full". Graphically (see figure 1), the vertical axis represents the economic welfare realm, measured through income and the horizontal axis represents the social rights realm, measured through the social privation index, which ranges from 1-6. This is, people located closer to the vertical axis have more shortcomings. Thus, quadrants are described as follows: **a) Multidimensional poor:** People under the economic welfare line and who have at least one social right violated; **b) Vulnerable by social rights:** People that present one or more social rights violated but their income is above the economic welfare threshold; **c) Income-related vulnerable:** People who have their social rights met but that are below the welfare threshold; and **d) Non-poor:** People above the welfare line and which have all rights met.

**Figure 1. Schematic representation of CONEVAL official multi-dimensional poverty measurement**

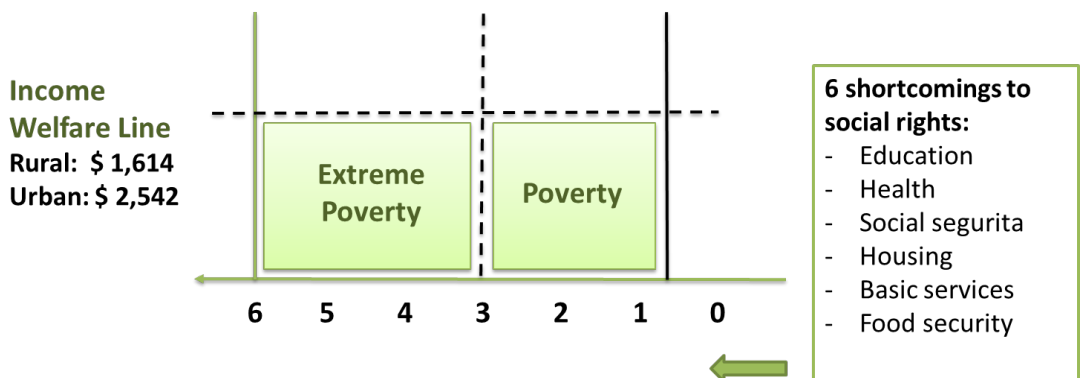

Source: Adjusted from CONEVAL

Thus, according to CONEVAL, a person is poor, multi-dimensionally speaking, if she has shortcomings both in terms of welfare and any of its social rights, with some additional measures relating to the deepness and unevenness in monetary poverty. In terms of indicators, let us use the current indicator of school attendance.

**School_att** ⎰ 1 if relevant household member attends any school within the National Education System

0 otherwise

Using this conceptual framework, Mexico measured poverty multi-dimensionally for the first time in 2008. This exercise has been repeated in 2010, 2012 and 2014.

## 2. Why including an environmental dimension?

The quantity and quality of ecosystems and the environment have at least three direct relationships with poverty. First, as provider of key goods and services for human survival such as food, water and fibers for isolated rural populations (Costanza et. al. 1997, 2007, 2009, 2014; Millenium Ecosystem Asessment, 2005). Second, healthy ecosystems result in higher outputs per unit of effort such as agricultural productivity derived from healthy soils (Palmer, C & Di Falco, S; 2012). And third, a fit environment results in better human capacities for productivity due to a good health derived from an adequate air or water quality, for example (Millenium Ecosystem Asessment, 2005).

United Nations has recently adopted the 17 global sustainability indicators which identify critical welfare-based environment elements: water, air, biodiversity, energy and waste, under three common criteria: i) effective access ii) stability and iii) quality.

**Water.**  In the last few years, it has been stressed in a variety of international scientific gatherings that water quality and availability is one of the top civilization challenges.  In 2000, the UN already forecasted that the global hydrological cycle would hardly be able to meet and adapt to the increasing global water demands (UNEP, 1999). The UN also recognizes that water is a centerpiece of sustainable development, affecting other critical sustainable development elements such as the effective access to energy and food (UN, 2015).

**Energy.** The UN General Assembly has declared 2014 to 2024, the "Energy For All" decade (UN, 2013), which underscores the need to support the 1.3 billion people who lack access to modern electricity and approximately 2.6 billion who depend on traditional energy sources to meet their heat and cooking needs (wood, animal manure, etc), with substantial health impacts. Also, there is increasing evidence that the poor spend a large proportion of their time and income satisfying their energy needs (see for example Falkingham & Namazie, 2002; Khandker et al, 2010; Kebir & Philipp, 2004; Bhuiya et al, 2007; Sovacool, 2012; UNDP, 2003; WHO, 2006).

**Biodiversity.** The UN Convention on Biological Diversity (CBD) defines biodiversity as "the variability of living organisms within terrestrial, aquatic and other ecological systems". The relationship between biodiversity and poverty has been understood from two fundamental premises: (i) as a source of livelihoods and/or income and/or (ii) biodiversity as a cost-effective insurance mechanism for food security and the protection against natural disasters (see CBD, 2010). Also, it has been noted that biodiversity has a higher relevance for households with fewer economic means (Adams et al. 2004; Palmer & Di Falco, 2012; Millenium Ecosystem Asessment, 2005) in particular vulnerable members within poor households (see for example Glaser, 2003), as well as for isolated communities (see for example Fu et al, 2009).

**Air.** According to key environmental indicators of the Organization for Economic Cooperation and Development (OECD, 2008), the quality of air and its influence in poverty is one of the major current global concerns. The World Health Organization (WHO, 2013) estimates that air pollution in interior spaces causes approximately 2 million premature deaths per year, causing almost half of the cases of pneumonia in children under 15 years. In outdoor spaces, atmospheric pollution causes 3.7 million premature deaths worldwide yearly. Air quality not only has impacts on health but also has a direct and indirect impact in education. Recently, the Center of Investigation in Environmental Epidemiology of Barcelona (Minguillón et al, 2015), demonstrated that kids that go to schools near traffic nodes show lower levels of cognitive development not only because of noise but because of air pollution. In particular, elemental carbon presence showed impacts of up to 13% in cognitive development.

**Noise, Mobility and Recreation.** Households´ welfare, particularly in the urban areas, can substantially be affected by spatial elements such as noise, mobility and the lack of spaces for recreation. In terms of mobility, the feasibility that a child gets to school or a parent to work, depends on an effective access to a conveyance, its continuity and the transfer time. Meanwhile, excessive noise is a factor that can alter the welfare and development opportunities of families that have few hours of sleep. In the most populated cities in the planet, mobility, noise and recreation are starting to be regarded as critical determinants of mental health and violence within households (Evans and English, 2002).

**Waste.** The inappropriate disposal of garbage and dangerous residues has been historically considered one of the important environmental hazards (Kahn & Kahn, 2009). With population growth and the increasing demand for food and other commodities, lots of waste is generated in households (also known as solid waste). The fact that waste is not properly collected and transferred to confined landfills can cause serious health problems for households (see for example Boadi and Kuitunen, 2005). Meanwhile, hazardous waste may contain chemicals, heavy metals and substances generated as by-products during commercial manufacturing processes, as well as household products such as paint thinner, cleaning fluids and old batteries. Without good waste management practices, chemical contents may pose a high risk to human health, especially in low-income populations.

**Climate Change.** Science tells us that we are tied to a temperature increase of at least 2 ° C and between 4 ° C and 6 ° C by the end of this century (IPCC, 2014). Climate change is expected to be one of the major aggravators of poverty in the coming years (WB, 2015). Households with low incomes are not only more vulnerable to impacts of climate change, because of the location and characteristics of housing, but also they have lower adaptive capacities due their lack of insurance mechanisms (Dercon, 1996, 2005, 2006; Zimmerman & Carter, 2003).

3. **How to include the environmental dimension into multidimensional poverty measurement**

Different avenues can be pursued to conceptualize the exercise to the right to a healthy environment from a rights-welfare perspective. Internationally, in their first forays, the links between environment and welfare were understood from the point of view of environmental health vis to vis the health of people, through what is called generically "environmental health". However, over the years, it has become increasingly evident that ecosystems' health and, the quantity and quality of the environmental services they produce such as water purification, plant pollination, pest control, climate regulation or protection to environmental hazards such as floods, landslides and hurricanes are critical to human wellbeing (Millennium Ecosystem Assessment, 2005).

**Figure 2. Basic environmental sub-dimensions for development and welfare**

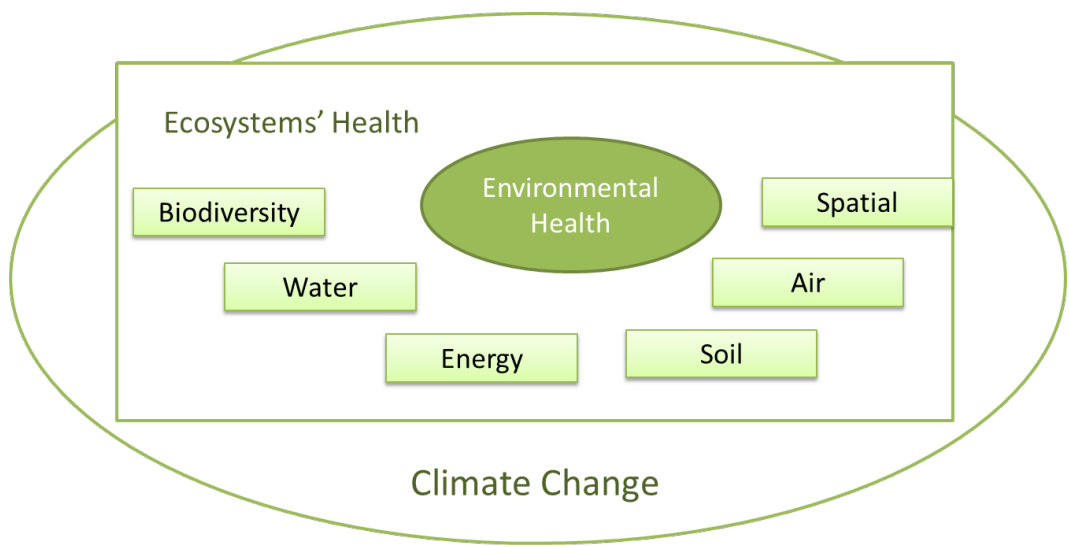

We thus propose two critical sub-dimensions. On the one hand, *Ecosystem Health*, understood as the quantity, quality and resilience of ecosystems and their critical goods and services for welfare and, on the other hand, *Environmental Health*, defined by the World Health Organization (WHO, 2013) as the impacts on human health from the interactions with their proximate environment. Climate change would play an "over-arching" catalysing external impact role, affecting the ability or efficiency of ecosystems to produce goods and services critical to human well-being, as well as potential direct impacts on people.

Sticking to the ground-rules set by CONEVAL, a deficiency in the environmental dimension of poverty should reflect a "pronounced deprivation in welfare", by shortcomings which individually or collectively, are essential for the survival of households and / or a decent life for its members.

The indicators to build the *Indicator of Wellbeing Deficiency from Environmental Poverty*, are grouped into two. The first including key ecosystem elements namely water, energy and biodiversity. And, the second, including the most important environmental health elements: air, spatial health and waste (see table 1).

**Table 1. Variables for the wellbeing deficiency due *Environmental Poverty***

| Ecosystem Poverty | Poverty in Environmental Health |
|---|---|
| 1. Water poverty | 4. Respiratory health |
| 2. Energy poverty | 5. Spatial health |
| 3. Biodiversity poverty | 6. Sanitation health |

Source: Own

Furthermore, we propose the incorporation of an *Indicator of Poverty Vulnerability due to Climate Change* that identifies those households whom are likely to become poor, given their extreme exposure and sensitivity to climate change events and, their low adaptive capacity.

## Ecosystem Poverty

1. **Water Poverty**

We understand water poverty as "deficiencies in one or more services, critically related to water supply,  resulting in not having secured access to sufficient good quality water to meet daily needs" (Subbaraman et al 2015), which according to the urban or rural environment include: *a) Drinking Water.* This is the most vital and inelastic water use. Depending on the weather, humans need between 1 and 5 liters of water per day; *b) Domestic water.* This category represents other vital uses of water, the most relevant being the water used for cooking and then for hygiene and laundry purposes; *c) Water used to grow crops.* This category of water use is for people who need more water than previously considered, either to grow the food they consume (or part of it), or to carry out other activities in which their livelihood depends critically. The most common example is that of small subsistence farmers; *d) Water for economic activities.* This refers to water uses that are part of the production of goods for people who are economically dependent on such production, but whose domestic food and basic needs are not affected drastically by scarcity; *e) Water for Ecosystems.* Since humans are part of the ecosystem, they are also affected when the quantity, quality and timing of water flows necessary to support ecosystems on which they directly depend. Considering the recently adopted global sustainable development goals, we take three generic criteria that will be used throughout the environmental elements proposed: *availability, quality and, effective access.*

Water poverty can thus be measured by an indicator composed of three sub-indicators:

*Water Availability.* Measured as the minimal quantity required for the members of a household to meet their basic needs[1]. Considering the minimum proposed by other healthcare organizations with volumes under 1.8 lt/day (EPA, 2000; Greenhalgh, 2001; Rome, 2010), we propose a minimal critical value of 2 lt/day per household member.

**Wat_avail**
- 1 if households has access to 2lt of water per day per member.
- 0 otherwise

*Effective Access.* Even if there is a sufficient volume of water available, water may be inaccessible due to lack of infrastructure or due to high costs relative to the household income or social conflicts that prevent access to water resources. In this regard, 20 percent of the poorest households in El Salvador and Nicaragua spend on average more than 10 percent of their income on water (UNDP, 2006), a level that we will take as a first threshold value.

**Wat_effacc**
- 1 if in order to have 2lt per household member per day must invest more than 10% of family income or it requires a long haul which translates into equivalent paid work of more than 10% of family income or if there is an intermittence in supply 50% or more of the time.
- 0 otherwise

With a certain volume of water ensured, a situation of water poverty can be generated by the bad quality of the resource. Four main pollutants are used as reference of bad water quality: (i) Biological Oxygen Demand (BOD) and (ii) Chemical Oxygen Demand (COD), which represent the pressure that a burden of pollution puts in a body of water. If this pressure or demand is too high, ecological processes within it will start to fail. For example, fish could not have sufficient oxygen available, causing mass mortality. A third pollutant, (iii) total suspended solids, groups several types of materials that according to their composition, can impact water users in different ways, subtracting clarity, requiring filtering, damaging infrastructure, having greater toxicity, etc. Finally, (iv) the concentration of faecal coliform is a pathogen that has a direct impact on health if ingested. Threshold values for these pollutants in Mexico are found in NOM-127-SSA1-1994. However, as in many other countries, Mexico is in lack of a complete monitoring system for water quality. Thus, following Puerto Rodríguez and collaborators (1999), we propose to use incidence of diarrheal diseases as a proxy for households in all municipalities lacking water monitoring or effective access.

---

[1] The amount of water required per person is difficult to calculate and more than a static figure, it is a dynamic value that depends on multiple factors such as gender, age, physical activity, diet and climate, among others. Some guidelines on a minimum dimension have proposed 2.9 L / day for an adult man of 70Kg and 2.2L / day for an adult woman of 58Kg. In both cases, intense physical activity in a hot environment raises requirements to 4.5L / day. Usually a child of 10 kg requires 1L / day and in the same way 4.5L / day under conditions of physical activity in hot environments (Grandjean, 2004; Howard & Bartram, 2003).

**Wat_qual** $\begin{cases} \text{1 if the household lives in a municipality where any of the NOM-127-SSA1-} \\ \text{1994 standards are violated or any member has reported periodic diarrheic} \\ \text{health events.} \\ \\ \text{0 otherwise} \end{cases}$

Then, an indicator of deficiency by Water Poverty would be defined as follows:

**WAT_POV_I** $\begin{cases} 1 & \textit{if Wat\_acc = 1 or Wat\_ effacc= 1 or Wat\_qual = 1} \\ 0 & \text{otherwise} \end{cases}$

**2. Energy Poverty**

In the last decade, there has been an honest attempt at defining alternative measures for energy poverty defined traditionally as the lack of effective access to modern energy sources (mainly electricity), without a consensus still (Nussbaumer et al, 2012). Three main approximations exist. The first is rooted in the definition of poverty as a minimal amount of food intake needed to sustain a healthy life (Barnes et al, 2011). Translating this approach to energy, it is assumed that there must be a minimum amount of energy needed to meet household needs. While this may be true, an adequate minimum may well depend on the type of food, culinary traditions, hours of light per day, weather, etc. The second approach defines it as the average energy used in households below the poverty line, defined by income (Pachauri et al, 2014; Barnes et al, 2010). This is quite attractive as it greatly simplifies measurement. However, this approach has the disadvantage to define energy poverty using indirect criteria, which means for example that energy poverty will not be directly linked with energy policies but with general economic and social policies, limiting its usefulness as a measure. The third line of thought is that energy poverty must be based on the percentage of household income spent on energy (Barnes et al, 2005). It is well established that poor households spend a higher percentage of their income on energy than rich households.  Several empirical studies indicate that the percentages may vary from about 5% to about 20% of income or cash expenses (WHO, 2006; UN, 2009). It is suggested that when energy spending is above 10% of the total household income, it is conceivable that it will begin to have an impact on overall household welfare. The idea is that when households are forced to spend as much as 10% of their cash income (Ten Percent Rule or TPR) on energy or the equivalent remuneration in time, they are being deprived of other goods and services necessary to maintain the basic life (Boardam, 1991;

Heindl & Schüssler, 2015). The TPR has been criticized for lacking scientific foundation and international comparability (Healy, 2004). Around the concept of affordability, several alternative proposals have been made (see Table 2).

**Table 2. Summary of energy affordability measurements**

| Measure | Energy poverty definition |
|---------|---------------------------|
| Ten percent measure (TPR) | Energy poor if the share of expenditure on energy relative to income exceeds ten percent. In the simulation study, the TPR is restricted to the poorest 30 percent of households. Households in income brackets above 30 percent are excluded by definition. |
| Two times median expenditure share (2M) | Energy poor if the expenditure share on energy exceeds two times the median expenditure share in the overall population |
| Low income high cost standard approach (LIHC) | Energy poor if household has expenditure on energy above the median and falls below the income poverty line after expenditure on all energy services |
| LIHC with TPR as first condition (LIHCt) | Energy poor if households has an expenditure share equal to or exceeding ten percent of income and falls below the income poverty line after expenditure on all energy services |
| LIHC with median expenditure of poorest 30 percent of households as first condition (LIHCm) | Energy poor if household has expenditure on energy above median expenditure of the poorest 30% of incomes and falls below the income poverty line after expenditure on all energy services |
| Minimum income standard (MIS) | Energy poor if disposable income after expenditure on all energy services is equal or less than the minimum income standard (MIS). |
| Quantile indicator (QI) | Energy poor if the expenditure share exceeds 'x-times some upper quantile of energy expenditure' (with $x < 1$) relative to income. |
| Income Poverty (RPL) | Energy poor if household falls below the (relative) income poverty line (RPL) after all expenditure on energy services. |

Source: Taken from Heindl & Schüssler, 2015 pg. 35

In a recent paper, Heindl & Schüssler (2015) compare these affordability approaches under some "dinamic" desirable behaviour: (i) a standard axiom in the literature is that poverty measurement should be monotonic (Sen, 1976); (ii) a measure of non-affordability increases if the actual expenditure on the respective goods increases in society without a change in individual positions. Based on these two criteria, the authors conclude that the LIHCt seems to be a good combination of the TPR and MIS and that it is also the only acceptable indicator in terms of its dynamic properties.

Using what we believe most valuable from these approaches, we propose three sub-indicators to measure energy poverty.

*Energy Availability.* Using the thresholds proposed by Tennakoon (Tennakoon, 2009) and the World Energy Outlook (IEA, 2015) we propose to establish a threshold value of 35 kg / member / year of liquefied gas or equivalent and 120 kWh / year of electricity.

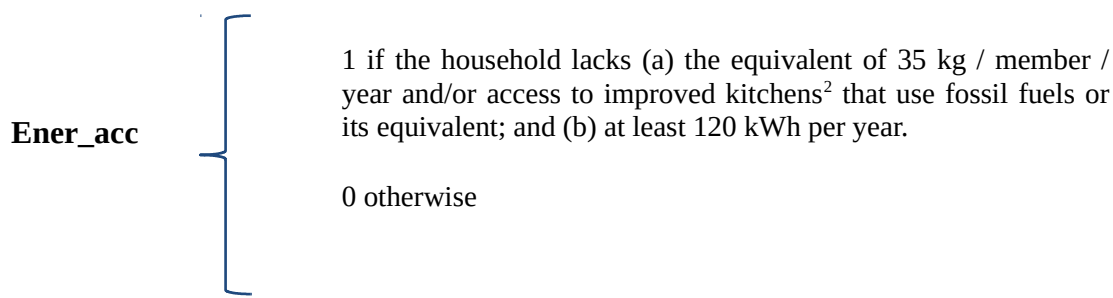

**Ener_acc**

1 if the household lacks (a) the equivalent of 35 kg / member / year and/or access to improved kitchens[2] that use fossil fuels or its equivalent; and (b) at least 120 kWh per year.

0 otherwise

*Effective access.* Measures the lack of access to electricity because of its high costs, or that accessing it, the cost deprives the home from a considerable amount of other goods and services due to the percentage of income that their acquisition represents.

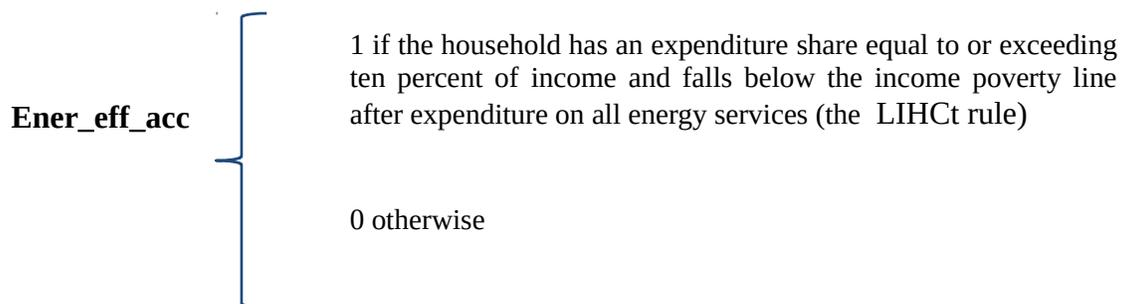

**Ener_eff_acc**

1 if the household has an expenditure share equal to or exceeding ten percent of income and falls below the income poverty line after expenditure on all energy services (the  LIHCt rule)

0 otherwise

*Quality.* It measures the quality of the service of energy distribution which could trigger a situation of energy poverty by having to take significant compensatory measures.

---

2 An improved wood stove is one that requires less than 4 person hours per week per home to get fuel, it fulfils the recommendations of the World Health Organization (WHO) for air quality (maximum concentration of CO of 30 mg/m$^3$ for periods of 8 hours of exposure), and efficiency is 25% or greater.

**Ener_qual** 〈 1 if in the household, the intermittence in electricity or gas availability, is greater than 50% of the time.

0 otherwise

*Energy Poverty Indicator*

**ENER_POV_I** 〈 *1 if Ener_acc = 1 or Ener_eff_acc = 1 or Ener_qual = 1*
*0 otherwise*

3. **Poverty in Biodiversity**

The Convention on Biological Diversity of the United Nations (CBD, 2010) defines biodiversity as "the variability among living organisms of all sources including, inter alia, terrestrial, marine and other aquatic ecosystems and the ecological complexes of which they are part; this includes diversity within species, between species and of ecosystems."

In the context of this paper, we will take biodiversity to include: (i) natural vegetation (forests, mangroves, and shrubberies); (ii) timber products; (iii) non-timber products; (iv) wild animals (including fish) and; (v) wild plants.

We propose three sub-indicators:

*Dependency*. Measures how dependent is a home directly on biodiversity.

**Biod_dep**
1 if income or household consumption depend 50% or more on biodiversity (fishing, timber, non-timber, wild animals or plants)[3].

0 otherwise

*Effective access*. Measures an important loss in the effective access to the resource.

**Biod_eff_acc**
1 if because of issues of land ownership, social conflicts, eviction or commons governance, the access to the resource has been lost at least 50% of the time.

0 otherwise

*Quality*. Measures an important loss in the volume or quality of the resource.

**Biod_qual**
1 if the volume or the quality of the resource has diminished in more than 50% over the last 3 years.

0 otherwise

*Poverty in Biodiversity Indicator*

**POV_BIO_I**
*1 if Biod_dep = 1 and, Biod_qual = 1 or Biod_eff_acc = 1*

0 otherwise

---

[3] In a recent global literature review conducted by the CBD on the relationship between poor households and biodiversity, it was found that biodiversity in rural households accounts for between 12% and 90% of income (CBD, 2010).

Using these three indicators, an indicator of **Poverty due Lack of Ecosystems's Health** is constructed:

**ECOS_HEA_POV**
$\Bigg\{$
*1 if POV_WAT_I = 1 or POV_ENER_I = 1 or POV_BIO_I = 1*

0 otherwise

## Environmental Health

### 4. Respiratory Health

Mexico has defined air quality standards for the main air pollutants ($SO_2$, $NO_2$, $PM_{10}$, $PM_{25}$, CO, $O_3$ y Pb), which are published in the Official Journal of the Federation (DOF). For example, the concentration of sulphur dioxide ($SO_2$) must not exceed 288 ug / $m^3$ or 0,110 ppm average in 24 hours, once a year. Table 3 shows the comparison between pollution standards in Mexico and standards defined by the World Health Organization in 2013. The first number listed is the international standard, while the second figure is the Mexican standard.

**Table 3. Comparison of national and international standards in air pollutants***

|          | Annual average | 24 hr average | 8hr average | 1 hr average |
|----------|----------------|---------------|-------------|--------------|
| $PM_{25}$ | 10,  40        |               |             |              |
| $PM_{10}$ | 20,  12        |               |             |              |
| $O_3$    |                |               | 100,  NE    |              |
| $NO_2$   | 40,  NE        |               |             | 200,  395    |
| $SO_2$   |                | 125,  288     |             |              |

*All numbers expressed in milligrams per cubic meter

In general, it can be observed that Mexican standards are higher. Yet, considering country particularities, it is suggested to use national standards as the reference, coupled with categories of vulnerability to health impacts, as officially defined (see Table 4).

**Table 4.  People in health vulnerability by social group (2000 and 2010)\***

|  | 2000 (Million) | 2011 (Million) |
|---|---|---|
| Old adults (60 yrs or more) | 7 | 10.5 |
| People with disabilities | 2.2 | 5.7 |
| Children under 5 yrs | 10.7 | 10.5 |
| Single mothers | 5.1 | 7.8 |
| People with HIV | 19,847 | 49,975 |

Source: Centro de Estudios de las Finanzas Públicas de la Cámara de Diputados con datos del INEGI; Anexo estadístico del Cuarto Informe de Gobierno.

In regards to indoor pollution, it is estimated that about 16.4 million people in Mexico´s rural households use wood as their main source of energy for heating (Masera, 1996). This, when coupled with a sole room and children, has been found to have extremely damaging effects (Edwards & Langpap, 2012).

We thus propose two sub-indicators to build the *Respiratory Health Indicato*r.

*Air quality in outdoor spaces.* It measures if the household is below a minimum welfare threshold, with respect to the main pollutants.

**Ext_air_qual**

1 if municipal air quality exceeds the average maximum within 24 hours, as nationally normed, once a year, for all pollutants ($NO_x$, $SO_2$, $PM_{10}$, $PM_{25}$ and CO) and the household head belongs to any of the health vulnerability groups described in Table 4.

0 otherwise

*Air quality in indoor spaces.* Measures if the household is below a wellbeing threshold due to constant exposure of vulnerable members to pollutants produced by directly burning fossil fuels.

**Ind_air_qual**  $\Big\{$  1 if the household uses firewood or other source of biomass, as the main source for heat, it does not have a separate room for cooking and children less than 15 years old live in the household.

0 otherwise

*Poverty due lack of Respiratory Health*

**POV_RESP_HEA_I**  $\Big\{$  *1 if Ext_air_qual = 1 or Ind_air_qual = 1*

0 otherwise

5. **Spatial Health**

We consider three elements to be critical to the spatial health of a household: mobility, noise, and recreation. In terms of mobility, we focus on finding a tolerable amount of time to get back and forth from work, school and / or the main economic activity of the household. As cities become larger, this is becoming a key issue for poor urban households as well. In Mexico City, for example, it is estimated that the minimum commuting time for a household back and forth is 30min while the longest is 6 hours (INEGI, 2007).

*Commute time.* Measures a threshold time for commuting to work or school.

**Comm_time**

1 if any member of the household spends more than 2 hours in an individual trip to get to work or school.

0 otherwise

In addition to transport times, access to transport is essential so that a home can access to work, schools and / or health services. An indicator of access to transportation to determine whether the home is at a maximum distance of any means of transport. The most common standards distance to bus stops or train stations has been set at 400 meters (El-Geneidy et al, 2014).

*Mobility for disadvantaged groups.* As a complementary indicator, access to transport services for disadvantaged groups is suggested..

**Hadicap_mov**

1 if the household has a member in a state of blindness, partial or total immobility and considers that it completely lacks mobility because a lack of signalling, adaptation of spaces and / or family support.

0 otherwise

With regard to recreation, in recent years, the European Union (EU) has been given the task of establishing a series of health indicators in recreation. For example, the EU states that any home at a distance of less than 500 meters from a park, green area, garden, museum or cultural centre is considered a home within an acceptable threshold for recreation. Given the focus of this work, we propose to focus on the access to green spaces. Threshold is selected using a 30min walk standard.

*Recreation.* Measures the minimum distance to a park or another green recreation area in which, combined with other indicators results in household welfare below a minimal theshold.

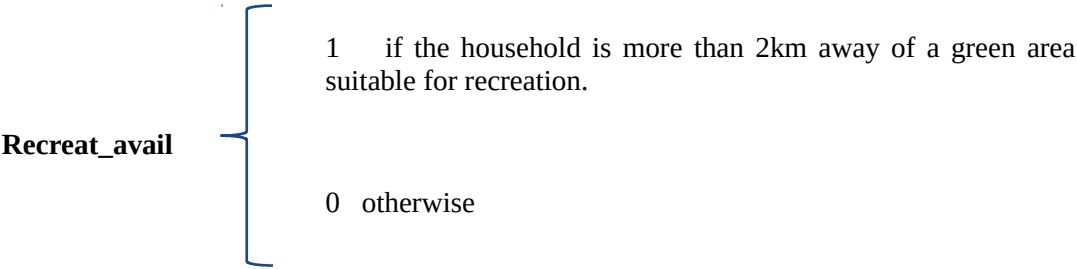

**Recreat_avail**

1    if the household is more than 2km away of a green area suitable for recreation.

0    otherwise

Meanwhile, the noise is a factor that can alter the welfare and performance of families in the minimum hours of sleep and be a determining factor for violence (see Liberzon et al, 2015; Evans et al, 2001). Also, recently published studies have found statistically significant relations between aircraft noise for example and a higher risk for strokes, coronary and cardiovascular diseases (see Huss et al, 2010; Correia et al, 2013).

*Noise at households.* Measures if the household is at an intolerable distance to some major source of noise.

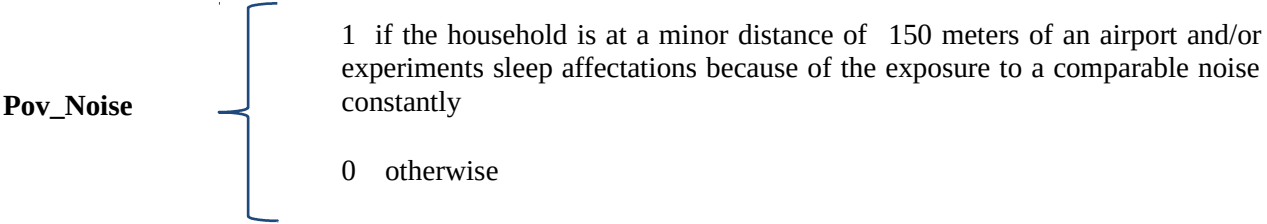

**Pov_Noise**

1  if the household is at a minor distance of  150 meters of an airport and/or experiments sleep affectations because of the exposure to a comparable noise constantly

0    otherwise

Based on indicators of mobility, leisure and noise described above, we define what we call Space Marginalization Indicator as follows:

*Space Marginalization Indicator*

**I_SPAT_MARG**

*1 if Comm_time = 1 or Handicap_mov = 1 or Recreat_avail = 1 and Pov_noise = 1*

*0 otherwise*

That is, we consider that a household is in a situation of spatial marginalization when it has mobility problems (one of the two sub-indicators of mobility) or has noise problems and at the same time a lack of access to recreation as a pronounced deprivation in well-being that can trigger negative mental or heart conditions and / or events of violence in the household.

6. **Health in Waste Management**

*Waste disposal.* Measures the main practice of waste disposal in households and the existence of direct harmful practices.

**Waste_disp**

1   if the household disposes the waste in a forest, the countryside, a lake or river

2   if the household disposes and/or burns the waste in its backyard, as main form of waste disposal.

3   if the household occupies a practice that represents more than 10% of its time or income as a main practice to dispose waste.

0   otherwise

*Closeness to open landfills.* If the home is located below a minimum distance to an open landfill. The Australian Environmental Protection Attorney has issued a guideline for minimum distances to several forms of pollutants and recommends households be located at least 200 meters away from a well-managed landfill (CIE, 2014). In terms of open landfills which continue to exist in most developing countries, a recent paper examining health effects in urban areas of Brazil, found direct health effects in distances below 1,000 meters (Gouveia & Ruscitto do Prado, 2010).

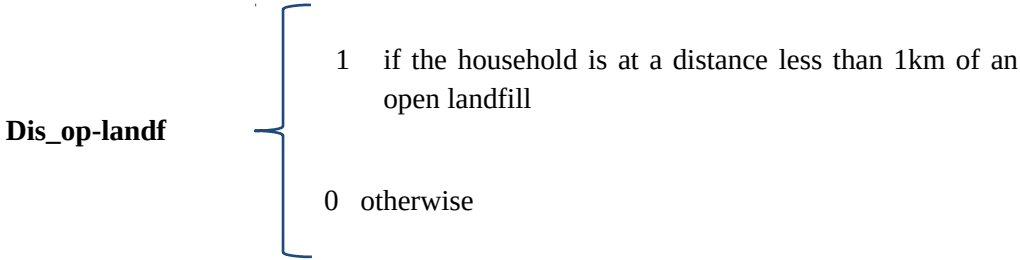

*Proximity to sites of improper disposal of hazardous waste.* Measures whether the home is exposed to mismanaged hazardous waste. The Australian Environmental Protection Attorney recommends a household distance of at least 1,000 meters for industrial chemicals (CIE, 2014).

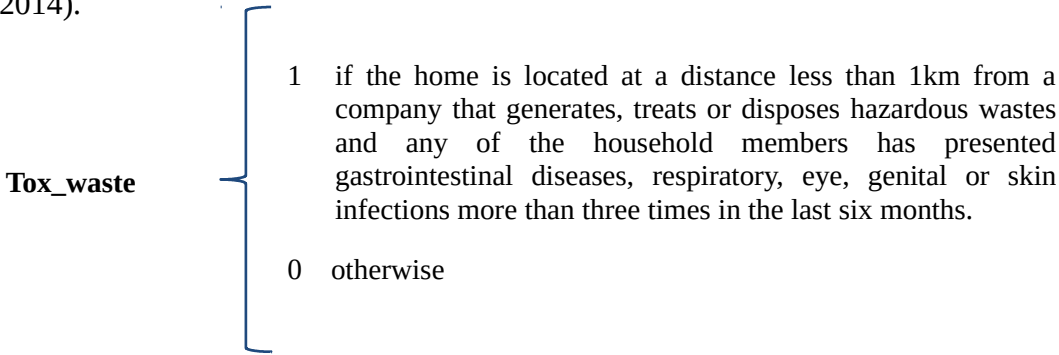

*Labour exposure to hazardous waste.* Not all households have members who may be exposed to hazardous materials. This sub-indicator pretends to measure the degree of exposure of the household head if engaged in activities related to mismanaged hazardous materials.

**Lab-tox-waste**

1    if the household head works in some highly polluting company[4] and has presented gastrointestinal diseases, respiratory, eye, genital or skin infections more than three times in the last six months.

0    otherwise

With the four indicators mentioned before, a *Poverty Indicator due Inadequate Waste Management* is constructed, as follows:

**I_POV_WASTE**

*1*    *If waste_disp=1 or dist_op_landf=1 or tox_waste=1 or lab_tox_waste = 1*

*0*    *otherwise*

Using the previous sub-indicators, a *Poverty due Lack of Environmental Health* indicator is constructed:

**POV_ENV_HEA**

*1*    *I_POV_RESP_HEA = 1 or I_SPAT_MARG= 1 OR I_POV_WASTE = 1*

0    otherwise

Finally, an indicator of shortcoming from Environmental Poverty is proposed:

---

4 Companies considered to produce toxic materials include agriculture, food industry, plastics, metals, asbestos, automotive, cellulose and paper, cement and lime, communications, freezing, construction, electronic equipment, exploitation of material banks , exploration and mining, forestry, power generation, wood, metallurgy, petroleum and petrochemical, paints and inks, clothing, chemical, steel and textile (CIE, 2014).

| | | |
|---|---|---|
| **A. SC_ENV_POV** | *1* | ECOS_HEA_POV = *1 or* POV_ENV_HEA = 1 |
| | 0 | otherwise |

**B. Vulnerability to Poverty from Climate Change Indicator**

According to the IPCC (2014), there are three factors that determine the high vulnerability of a household to climate change: *i) exposure,* which identifies the degree to which a system is exposed to important climate variations, *ii) sensitivity,* understood as the level at which a system would be affected, either negatively or positively, by climate-related stimuli; and *iii) adaptation capacities,* as the capacity for adjustment in response to actual or expected climatic stimuli or their effects, which moderates harm or exploits beneficial opportunities.

In the context of this paper, it seems essential to identify those households that, according to multidimensional poverty measurements can be considered not poor but due to their high exposure to extreme hydro-meteorological phenomena (drought , flood, hurricane or heat wave), their high sensitivity to suffer extreme impacts and their low responsiveness (adaptation), are at the brink of becoming poor. Using the schematic representation of CONEVAL's methodology depicted in figure 1, these households could be located just besides the "poverty quadrants", at the right side of the squares.

*Exposure.* Measures how exposed is a household to hydro-climatic phenomena.

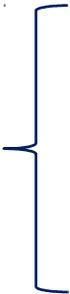

**Exp_cc**

if the household belongs to a municipality classified by the INECC as:
1  Highly exposed to droughts
2  floods
3  hurricanes and/or
4  heat waves

0  otherwise

*Sensitivity.* Measures how sensitive a household to be impacted by a hydro-climatic phenomenon given its location or the materials of which is built from.

**Sens_cc**

1  if Exp_cc = 1 and the income or household consumption depends 50% or more on own or common property land and more than 50% of the land is not irrigated;
2  if Exp_cc= 2 and the household is located on a mountain slope or river channel;
3  if Exp_cc= 3 and the household has no ventilation and has elder members with health problems; and
4  if Exp_cc= 4 and the household lacks proper housing materials as defined by CONEVAL.

0  otherwise

*Adaptive capacity.* Measures how likely it is for a household to recover from climate impacts.

**Adap_cc_cap**

1  if the household has savings for at least 3 months of its average monthly income or has secured access to a credit of similar amount.

0 otherwise

With the three indicators described above, an *Indicator of High Vulnerability to Poverty due to Climate Change* is constructed, as follows:

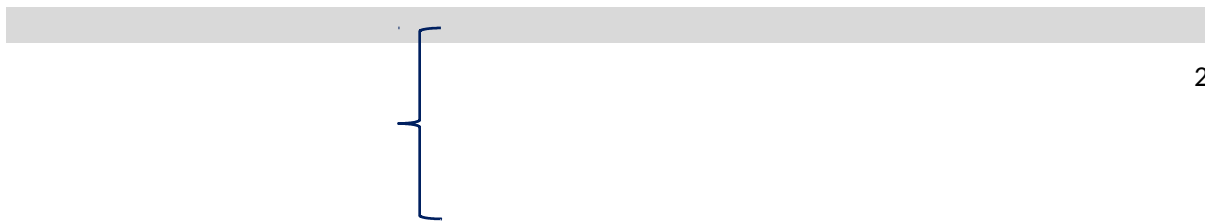

| VUL_POB_CC | $1$ | if Exp_cc = 1, 2, 3 or 4 and Sens_cc = 1, 2 3 or 4 and Adap_cc_cap= 1 |
| | 1 | otherwise |

**Figures and Tables**

Figure 1. Schematic representation of CONEVAL official multi-dimensional poverty measurement

Figure 2. Basic environmental sub-dimensions for development and welfare

Table 1. Variables for the wellbeing deficiency due *Environmental Poverty*

Table 2. Summary of energy affordability measurements

Table 3. Comparison of national and international standards in air pollutants

Table 4. People in health vulnerability by social group (2000 and 2010)